\documentstyle[12pt,psfig]{article}

\newcommand{\mathbold}[1]{\mbox{\protect\boldmath $\displaystyle #1$}}
\begin{document}

\title{$\rho$-meson properties at finite nuclear
density\thanks{Supported by DFG, Forschungszentrum J\"ulich
and BMBF.}}

\author{L.A. Kondratyuk$^1$ , A. Sibirtsev$^2$, W. Cassing$^2$ \\ 
Ye.S. Golubeva$^3$ and M. Effenberger$^2$ \\
$^1$ Institute of Theoretical and Experimental Physics, \\
117259 Moscow, Russia\\
$^2$ Institute for Theoretical Physics, \\
University of Giessen, D-35392 Giessen, Germany \\
$^3$ Institute of Nuclear Research, \\
117312 Moscow, Russia}
\date{ }
\maketitle

\begin{abstract}
We calculate the momentum dependence of the $\rho$-meson
selfenergy based on the dispersion relation for the $\rho N$
scattering amplitude $f(\omega$) at low nuclear density. The
imaginary part of $f(\omega)$ is determined from the optical
theorem, while the total $\rho N$ cross section is obtained
within the VDM at high energy and within the resonance model at
low energy. Our numerical results indicate a sizeable broadening
of the $\rho$-meson width in the medium especially for low
relative momenta $p$ while the real part of the $\rho$
selfenergy is found to change its sign and to become repulsive
already at momenta above 100 MeV/c. Extrapolating to 
nuclear saturation density $\rho_0$ we
find a dropping of the $\rho$-mass for $p \approx$ 0 roughly in line with
the QCD sumrule analysis of Hatsuda while at high energy an
increase of the $\rho$-mass close to the prediction by
Eletsky and Joffe is obtained. However, when including a broadening
of the baryonic resonances in the medium, 
the $\rho$-meson mass shift at $p \approx$ 0 becomes slightly repulsive
whereas the width increases substantially.
\end{abstract}
\vspace{1cm}


\newpage
\section{Introduction}
The properties of baryonic and mesonic resonances in the nuclear
medium has received a vivid attention during the last years (cf.
Refs.~\cite{Brown,Hatsuda,Asakawa,Shakin,Klingl}) 
within the studies on the
properties of hot and dense nuclear matter. Here, QCD inspired
effective Lagrangian models~\cite{Brown,Shakin,Klingl} or approaches
based on QCD sum rules~\cite{Hatsuda,Asakawa} predict that the masses
of the vector mesons $\rho$ and $\omega$ should decrease with
the nuclear density. On the other hand, with a dropping hadron
mass the phase space for its decay decreases, which results
in a modification of its width or lifetime in matter, while due
to interactions with the surrounding nuclear medium the
resonance width will 
increase ~\cite{Kondratyuk,Boreskov,Cassing,Golubeva}.

The in-medium properties of vector mesons have been addressed
experimentally so far by dilepton measurements at the SPS, both
for proton-nucleus and nucleus-nucleus 
collisions~\cite{CERES,HELIOS,HELI,Drees}.
As proposed by Li {\it et al.}~\cite{Li}, the enhancement in $S+Au$
reactions compared to $p + Au$ collisions in the invariant
mass range $0.3 \leq M \leq 0.7$ GeV might be due to a shift of
the $\rho$ meson mass.  The microscopic transport studies in
Refs.~\cite{Cassing2,Cassing3,Brat} for these systems point in the same
direction, however, also more conventional selfenergy effects
cannot be ruled out at the present 
stage~\cite{Cassing2,Rapp,Wambach,Friman}.
Especially the p-wave coupling of the $\rho$-meson to nucleons
induces an attractive interaction at low relative 
momenta~\cite{Friman} which turns
repulsive at large momenta. The explicit momentum dependence of
the $\rho$-meson selfenergy thus is an important aspect to be
investigated both, theoretically and experimentally e.g. by
$\pi^- \ A$ reactions~\cite{Cassing,Golubeva,Schoen,Weidmann}.

The main goal of this paper is to calculate the explicit
momentum dependence of the $\rho$-meson potential at finite
nuclear densities within a dispersive approach that is based on
the resonance model at low relative momenta of the $\rho$ with
respect to the nucleon at rest and the vector dominance model at
high relative momenta following the suggestion by Eletsky and
Joffe~\cite{Eletsky}. Our work is organized as follows: In Section 2
we briefly recall the relation between hadron self energies and the
hadron-nucleon scattering amplitude at low density and discuss the
dominant interaction mechanisms. In Section 3  we present 
the related dispersion 
relations and evaluate the $\rho N$ scattering amplitude from the $\rho$-
photoproduction cross section at high energy. At 
low energy we use a resonance
model to determine the $\rho-N$ cross section.
The implications for the real and imaginary
part of the $\rho$-meson selfenergy at finite nuclear density 
are presented in Section 4 while a summary and
discussion of open problems concludes this work in Section 5.  

\section{Hadronic resonances in the nuclear medium}

The relativistic form of the wave equation  
describing the propagation of a mesonic resonance $R$ in the
nuclear matter is given as (see e.g.~\cite{Golubeva})
\begin{equation}
\label{waveeq}
\lbrack - \mathbold{\nabla}^2 + M^2_R - iM_R\Gamma_R 
+ U(\mathbold{r}) \rbrack  \ \Psi(\mathbold{r})=
E^2 \ \Psi(\mathbold {r}),
\end{equation}
where  $E^2 = \mathbold{p}^2+ M^2_R$ and 
$\mathbold{p} $, $M_R$  and $\Gamma_R$ are the momentum,
mass and width of the resonance, respectively.
The optical potential then is defined as
\begin{equation}
\label{optpot}
U(\mathbold{r})= - 4\pi \ f_{RN}(0) \ \rho_N(\mathbold{r})
\end{equation}
where $f_{RN}(0)$ is the forward $RN$-scattering amplitude 
and $\rho_N$ is the nuclear density.

It is useful to rewrite Eq.~(\ref{waveeq}) in the form
\begin{equation}
\label{waveeq2}
\lbrack \mathbold{\nabla}^2 + \mathbold{p}^2 \rbrack \ 
\Psi(\mathbold{r})  = \lbrack U(\mathbold{r}) -
 \Delta \rbrack \ \Psi(\mathbold{r}),
\end{equation}
where
\begin{equation}
\label{delta}
\Delta = P^2- M^2_R - iM_R \Gamma_R
\end{equation}
is the inverse  resonance propagator in the vacuum.
The four-momentum $P$ in Eq.~(\ref{delta}) can be defined 
through the four-momenta of the resonance decay products, i.e.
\begin{equation}
P=p_1+p_2+...
\end{equation}

When the resonance decays inside the nucleus of radius $R_A$
at density $\rho_N$, the propagator has the form
\begin{equation}
\Delta^*=\Delta+4\pi  f(0)\rho_N = P^2-M^{*2}_R+iM_R^*\Gamma_R^*
\end{equation}
where
\begin{equation}
\label{rmassa}
M_R^{*2}=M^2_R - 4\pi  Re{f(0)}  \rho_N,
\end{equation}
\begin{equation}
\label{rwidtha}
M_R^*\Gamma^*_R = M_R\Gamma_R + 4\pi Im{f(0)} \rho_N.
\end{equation}
Its spectral function then can be 
described (in a first approximation) by a Breit--Wigner formula
(nonrelativistically) as
\begin{equation}
\label{BW}
F(M)= \frac{1}{2 \pi} \ \frac{\Gamma^*_R}{(M-M^*_R)^2 +
\Gamma^{*2}_R/4} ,
\end{equation}
which contains the effects of collisional broadening,
\begin{equation}
\label{gammas}
\Gamma^*_R=\Gamma_R+\delta \Gamma 
\end{equation}
with
\begin{equation}
\label{dgamma}
\delta\Gamma = \gamma v \sigma_{RN} \rho_N,
\end{equation}
and a shift of the meson mass
\begin{equation}
\label{mstar}
M^{*}_R=M_R+\delta M_R 
\end{equation}
with
\begin{equation}
\label{dmstar}
\delta M_R=-\gamma v \sigma_{RN} \rho_N \alpha.
\end{equation}
In Eqs.~(\ref{dgamma}),(\ref{dmstar}) $v$ is the average resonance
velocity with respect to the target at rest, $\gamma$ is the
associated Lorentz factor, $\rho_N$ is the nuclear density while
$\sigma_{RN}$ is the resonance-nucleon total cross section and
$\alpha = Re f(0)/Im f(0).$

The sign of the resonance mass shift depends on the sign of the real part
of the forward $RN$ scattering amplitude which again
depends on the momentum of the resonance. For example, at low
momenta (p $\approx$ 0) various 
authors~\cite{Brown,Hatsuda,Asakawa,Shakin} 
predict a decreasing mass
of the vector mesons $\rho$, $\omega$ and $\phi$ with the
nucleon density, whereas Eletsky and Ioffe have argued 
recently~\cite{Eletsky} that the $\rho$-meson should
become heavier in nuclear matter at momenta of 2-7~GeV/c.  If
the ratio $\alpha$ is small - which is actually the case for the
reactions considered because many reaction channels are open -
the broadening of the resonance will be the main effect. As it
was shown in Ref.~\cite{Mosel} the account of this effect can
also essentially influence the predictions of QCD sum 
rules~\cite{Hatsuda,Asakawa}.
Therefore it is useful to perform independent calculations for the
real part of the $\rho N$ forward scattering amplitude which can
also be used at low momenta.

Whereas the $\rho$ -meson
spectral function in the nuclear medium has been evaluated in Refs.
\cite{Klingl,Rapp,Wambach} in dynamical models by considering its 
$2 \pi$ decay mode and taking into account the rescattering of pions on
nucleons in the medium, we here address an approach based on dispersion
relations using experimental information on the vacuum scattering
amplitude as input as well as on the $\rho_N$ couplings 
(cf. Ref.~\cite{Sibirtsev}).  Another important point is that 
Eqs.~(\ref{dgamma}) and (\ref{dmstar}) for the collisional 
broadening and mass shift
are valid only at low densities when the resonance-nucleon
scattering amplitude inside the nucleus is the same as in the vacuum.
A discussion of this point will be presented in Section~3.2 explicitly.

\section{The $\rho$-nucleon scattering amplitude in vacuum}

Within the framework of the Vector Dominance Model (VDM) the Compton
scattering amplitude can be expressed through the $\rho
N$, $\omega N$ and $\phi N$ scattering amplitudes as 
\begin{equation}
\label{coupl}
T_{\gamma N}(s,t) = \frac{e^2}{4 \gamma_{\rho}^2} 
\left\lbrack T_{\rho
N}(s,t) + \frac{\gamma_{\rho}^2}{\gamma_{\omega}^2}
T_{\omega N}(s,t)+ \frac{\gamma_{\rho}^2}{\gamma_{\phi}^2}
T_{\phi N}(s,t) \right\rbrack ,
\end{equation}
where $T_{V N}(s,t)$ is the invariant amplitude for the elastic
scattering of the transverse polarized vector meson on the nucleon,
$e^2 /4 \pi$ is the fine-structure constant, $T_{\gamma N}(s,t)$
is the invariant Compton scattering amplitude and 
${\gamma}_{\rho}^2/4\pi$=0.55~\cite{Anderson,Ballam,Egloff}.
According to experimental data (cf.~\cite{Bauer})
the last 2 terms on the r.h.s. of Eq.~(\ref{coupl}) including the
$\omega N$ and $\phi N$ amplitudes can
be neglected as compared to the first term. Then the $\rho N$
scattering amplitude can be expressed directly through the
Compton amplitude by multiplying the latter with $4 \gamma^2_\rho/e^2$;
this approach has been adopted by Eletsky and Ioffe in 
Ref. \cite{Eletsky}.
We note, however, that the experimental Compton scattering amplitude also
contains contributions from the continuum of $2 \pi$ and $n \pi$ 
intermediate states at high energy such that Eq. (\ref{coupl}) appears
questionable. Here we will adopt the experimental results on 
$\rho$-photoproduction instead which are more closely related to the
$\rho$-meson itself (cf. Section 3.1).  

In general, the $\rho N$ scattering amplitude 
$f(\omega,\theta)$, which enters into
Eqs.~(\ref{rmassa}),(\ref{rwidtha}) for the resonance mass 
shift and collisional broadening can be expressed through 
the invariant scattering amplitude by
\begin{equation}
\label{E1}
T_{\rho N}(s,t)= \ 8 \ \pi \ \sqrt{s} \ \frac{p_{cm}}{p_{lab}} \
f(\omega,\theta) ,
\end{equation}
where $p_{cm}$, $p_{lab}$ are the momenta of the incident
particle in the c.m. and laboratory systems,
respectively, while $\theta$ is the scattering angle in the
laboratory frame.  Furthermore,
the real part of the forward $\rho N$ scattering amplitude is
related to its imaginary part through the dispersion
relation (cf. Ref.~\cite{Bjorken})
\begin{equation}
\label{dis1}
Re f(\omega) = Re f({\omega}_0) +
\frac{2({\omega}^2-{\omega}_0^2)} {{\pi}} \
P \int_{{\omega}_{min}}^{+\infty} \
\frac{d{\omega}^{\prime} \  {\omega}^{\prime} \ Im f({\omega}^{\prime})}
{({\omega}^{\prime 2}-{\omega}_0^2)({\omega}^{\prime
2}-{\omega}^2)} ,
\end{equation}
where ${\omega}={\omega}_0$ is a substraction point and
${\omega}_{min}$ is the threshold energy.  The imaginary part of
the forward scattering amplitude $Im f(\omega)$, furthermore, is
related to the total cross section by the optical theorem,
\begin{equation}
\label{optic}
Im f(\omega)= \frac{p_{lab}}{4\pi} \ {\sigma}_{tot}(\omega).
\end{equation}
Thus the knowledge of the total cross section
$\sigma_{tot}(\omega)$ is sufficient to determine $Re f(\omega)$
through the dispersion relation~(\ref{dis1}) once the amplitude is known
'experimentally' at the subtraction point $\omega_0$.

\subsection{The $\rho N$ total cross section from photoproduction}

Within the VDM one can express the $\rho N$ scattering amplitude
not only through the Compton scattering amplitude (Eq.
(\ref{coupl})) but also through the amplitude 
for $\rho$- photoproduction as:
\begin{equation}
\label{gamrho}
T_{\rho N}(s,t) = {\frac{2 \gamma_{\rho}}{e}} \ T_{\gamma N \to
\rho N}(s,t). 
\end{equation}
Furthermore, the $\rho N$ total cross section can be related to
the differential cross section of the reaction $\gamma p \to
\rho p$ as 
\begin{equation}
\label{vector}
{\sigma}^2_{ \rho p} = {\frac{{\gamma}^2_{\rho}}{4\pi}} 
{\frac{64 \pi}{\alpha}} \ \frac{1}{1+ \alpha_{\rho p}^2} \
{\left(\frac{q_{\gamma}}{q_{\rho}} \right)^2} 
{\left. \frac{d\sigma_{\gamma p \to \rho p}} 
{dt} \right|_{t=0} }  .
\end{equation}
Here $q_{\gamma}$ and $q_{\rho}$ are the c.m. momenta
of the $\gamma N$ and $\rho N$ systems at the same invariant
collision energy $\sqrt{s}$. Furthermore, we assume that the ratio
$\alpha_{\rho p}$ of the real to imaginary part of the 
$\rho N$ forward scattering
amplitude is small. This assumption is valid at least for
energies above 3~GeV~\cite{Bauer}. Thus using the
$\rho$ photoproduction data from Refs.~\cite{Anderson,Ballam,Egloff} 
the $\rho N$ scattering amplitude is fixed at high energy.

We note that at higher energies
one can calculate the total cross section of the $\rho N$ interaction
also within the Quark Model (QM), where
$\sigma_{\rho N}$ can be expressed in terms of pion-nucleon
cross sections as
\begin{equation}
\label{QCD1}
{\sigma}_{ {\rho}^0 N}=\frac{1}{2} \left( {\sigma}_{ {\pi}^- N}
+ {\sigma}_{ {\pi}^+ N} \right) ,
\end{equation}
while  the $\pi N$ cross sections can be taken from a Regge fit to the
experimental data~\cite{Donnachie}. Though 
the additive quark model at first
sight appears questionable for Goldstone bosons as the pions, the cross
section $\sigma_{\rho N}$ will turn out practically 
the same as in the VDM 
for $\rho$-meson momenta above 2 GeV/c (see below).

\subsection{The resonance model combined with VDM}

If the VDM would be valid at all energies  
we could calculate the mass shift and collisional broadening of 
the $\rho$ meson in nuclear matter at low densities by inserting
Eq.~(\ref{gamrho}) in Eqs.~(\ref{rmassa}) 
and (\ref{rwidtha}) from Sect. 2. 
However, the VDM is expected to
hold only at high energies $\omega > 2$~GeV~\cite{Ioffe1} since
at lower energies ($\omega \leq $ 1-1.5 GeV)
there are a lot of baryonic resonances which couple 
strongly to the transverse as well as to the longitudinal $\rho
$-mesons~\cite{PDG}.  Therefore at low energies it is necessary to
describe the $\rho N$ interaction within the framework of a
resonance model~\cite{Sibirtsev,Martin}.
In the following we consider the $\rho N$ forward scattering
amplitude being averaged over the $\rho $-meson transverse and 
longitudinal polarizations.

Experimental information on the baryonic resonances and their coupling to
the $\rho$-meson is available for masses below 2 GeV.
We saturate the $\rho N$ total cross section at low energies by
the resonances listed in Table~1. For the Breit-Wigner
contribution of each resonance we adopt the approach developed
by Manley and Saleski~\cite{Manley}. In this model the total 
$\rho N$ cross section is given  as a function of the 
$\rho$-meson mass $m$ and the invariant collision energy
$\sqrt{s}$ as:
\begin{equation}
\label{resxsection}
\sigma_{\rho N}(m, \sqrt{s}) = \frac{2\pi}{3q^2_{\rho} }
\sum_R (2J_R+1)
\frac{s {\Gamma}_{\rho N}^{in}(m, \sqrt{s}) {\Gamma}_{tot}(\sqrt{s})}
{(s-M_R^2)^2 + s{\Gamma}_{tot}^2(\sqrt{s})}.
\end{equation}
Here $q_{\rho}$ denotes the c.m.  momentum of the $\rho N$
system, $J_R$ the spin of the resonance, $M_R$ the pole mass and
${\Gamma}_{tot}$ the total width as a sum over the partial channels.
For the case of an unstable particle in the final channel the
partial width has to be integrated over the spectral function 
of this particle. The energy dependence of
the partial width ${\Gamma}_{\rho N}$ for the decay of each
baryonic resonance into the $\rho N$ channel, i.e.  
$R \to \rho N$, is given by
\begin{equation}
\label{width}
{\Gamma}_{\rho N}(\sqrt{s}) \ = \ {\Gamma}_{\rho N}(M_R) \
\frac{g(\sqrt{s})}{g(M_R)} ,
\end{equation}
where ${\Gamma}_{\rho N}(M_R) $ is the $\rho N$ 
partial width at the resonance pole $M_R$, while
the function $g(\sqrt{s})$ is determined as
\begin{equation}
\label{enfun}
g(\sqrt{s}) = \int_{2 m_{\pi}}^{\sqrt{s}-m_N} A_{\rho}(m') \
\frac{q_{\rho N }(m')}{\sqrt{s}} \
B^2_l(q_{\rho N }) \ dm' .
\end{equation}
In Eq. (\ref{enfun})  $q_{\rho N}(m')$ is the c.m.  momentum of the
nucleon and the $\rho $-meson with mass $m'$,
$B_l$ is a Blatt--Weisskopf barrier penetration
factor, $l$ denotes the angular momentum of the $\rho N$ system 
and $A_{\rho}$ is the spectral function of the $\rho$-meson
in free space taken as
\begin{equation}
\label{spef}
A_{\rho}(m) = \frac{2}{\pi} \ \frac{m^2 {\Gamma}_{\rho}(m)}
{(m^2-M_{\rho}^2)^2+m^2 {\Gamma}_{\rho}^2(m)},
\end{equation}
where $M_{\rho}$=770~MeV and the mass dependent width
${\Gamma}_{\rho}(m)$ of the $\rho$-meson is given by
\begin{equation} 
{\Gamma}_{\rho}(m) =\frac{{\Gamma}_{\rho}^0 M_{\rho}}{m}
\left\lbrack \frac{{q_{\pi \pi }(m)}}{q_{\pi \pi }(M_{\rho})} 
\right\rbrack^3
{\left\lbrack \frac{1+{\delta}^2{q^2_{\pi \pi }(M_{\rho})}}
{1+{\delta}^2{q^2_{\pi \pi }(m)}} \right\rbrack}^2
\end{equation}
with $\delta =5.3$~(GeV/c)$^{-1}$ and
${\Gamma}_{\rho}^0=150$~MeV.

The incoming width in Eq.~(\ref{resxsection}) reads:
\begin{equation}
\Gamma_{\rho N}^{in}(m, \sqrt{s} )=
C_{\rho N}^{I_R} \frac{q_{\rho N}(m)}{\sqrt{s}} B_l^2(q_{\rho N})
\frac{\Gamma_{\rho N} (M_R)}{g(M_R)},
\end{equation}
where $ C_{\rho N}$ denotes the appropriate Clebsh-Gordan
coefficient for the coupling of the isospins of $\rho $ and 
nucleon to the isospin $I_R$ of the resonance. 
The properties of the baryonic resonances coupled to 
the $\rho$ are listed in Table~1; in the following 
calculations we exclude the resonances with 
only one star confidence level.

\begin{figure}[h]
\psfig{file=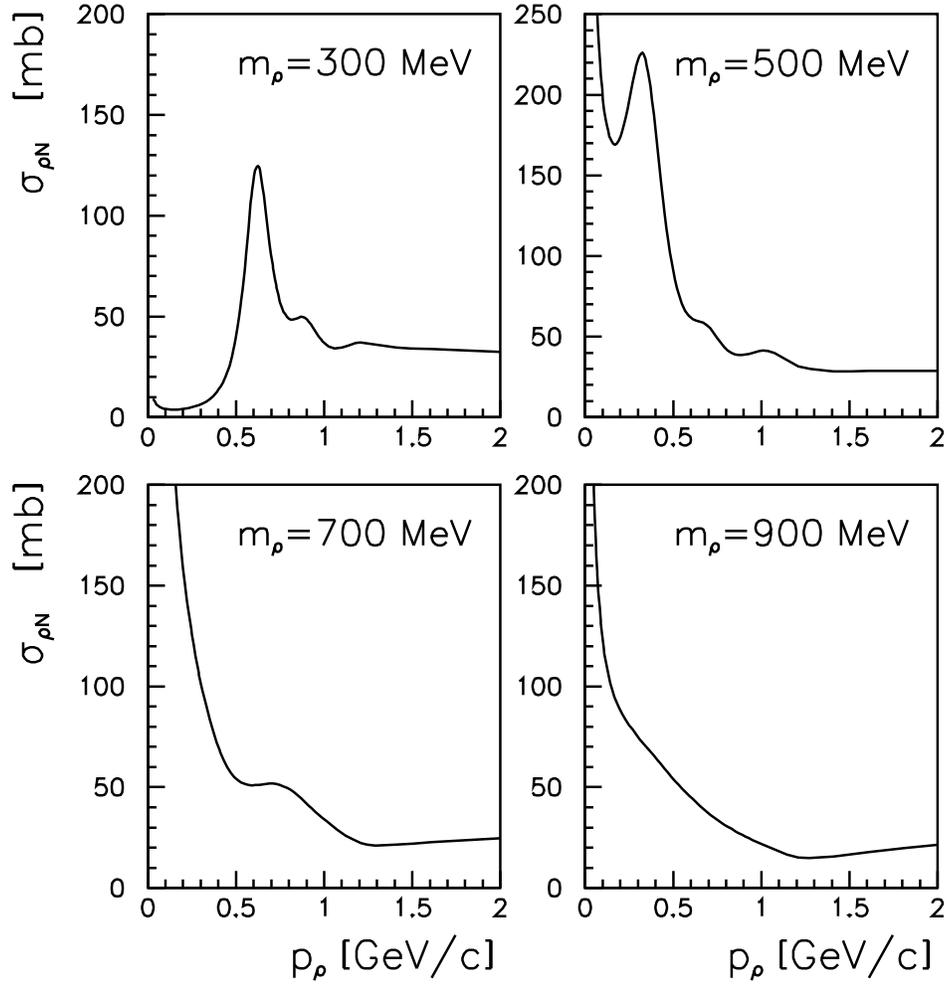,width=15cm}
\caption[]{\label{lk15} The total  $\rho N$ cross section
for different invariant masses of the $\rho $-meson. At low energy 
the cross section was obtained within the resonance model while
at high energies it was extrapolated from the quark model (QM) 
(\protect\ref{QCD1}).}
\end{figure}

Within the resonance model we can evaluate
the total $\rho N$ cross section as the function of two
variables: the $\rho$ momentum and the invariant 
mass $m$ of the $\rho $-meson.
Fig.~\ref{lk15} shows the momentum dependence of the total
$\rho N$ cross section according to (\ref{resxsection}) for different 
masses of  the $\rho $-meson, respectively. At higher momenta
($p_\rho \ge 1.5$~GeV/c), where the
total cross section is described by the VDM or the QM (\ref{QCD1}), 
respectively, we assume that it does no longer depend on $m$. 

\begin{figure}[h]
\psfig{file=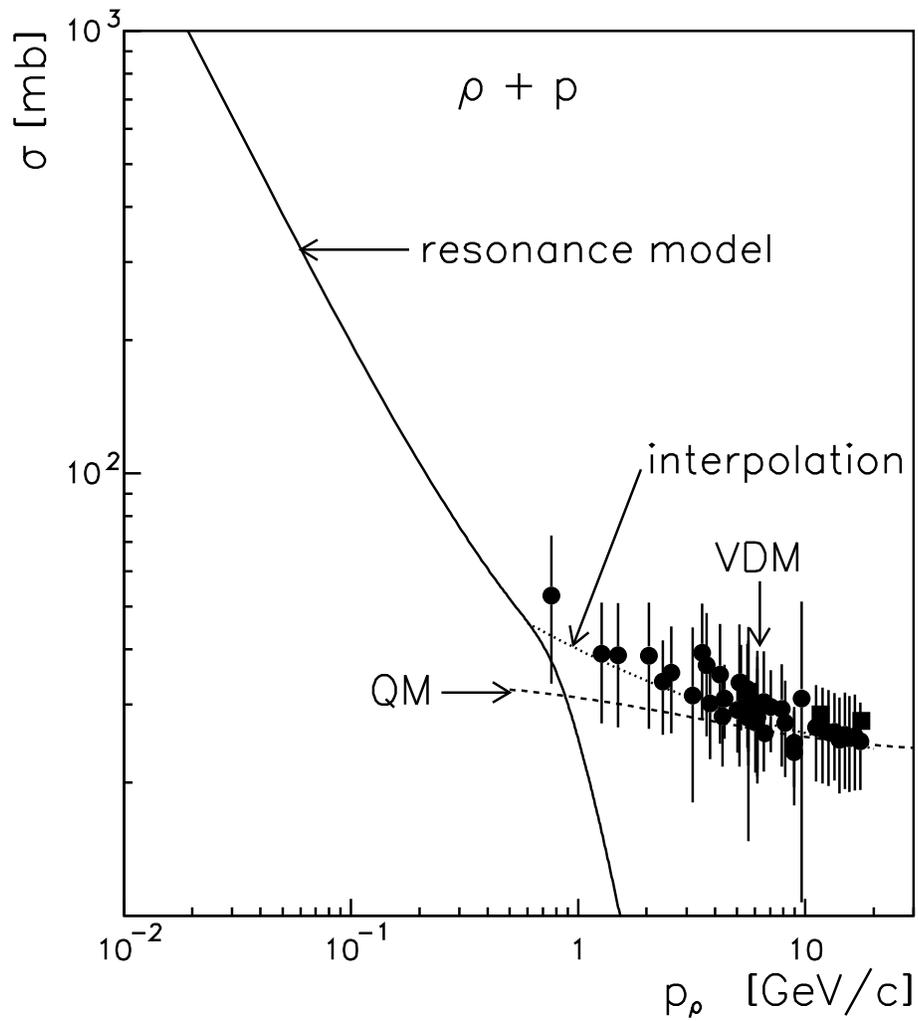,width=15cm}
\caption[]{\label{lk5} The total  $\rho p$ cross section.
The solid line shows our calculation within the resonance model 
(\protect\ref{resxsection}) while
the dashed line is the result from the quark model (QM)
(\protect\ref{QCD1}). The full
circles show the experimental data extracted from 
$\rho$-photoproduction while the squares are
from~\protect\cite{Anderson1}. The dotted line indicates the
interpolation that will be used furtheron.}
\end{figure}

In Fig.~\ref{lk5} we present the prediction of
the resonance model for the total $\rho N$ cross section --
averaged over the $\rho$-meson spectral function 
$A_{\rho}$ -- in comparison 
to the result from the quark-model (\ref{QCD1}) (QM, dashed line).
The solid circles in Fig.~\ref{lk5} show the total $\rho p$
cross section obtained by Eq.~(\ref{vector}) and the
experimental forward $\rho$-photoproduction data
from~\cite{Anderson,Ballam,Egloff}. Furthermore, the squares in
Fig.~\ref{lk5} (for $p_\rho \geq$ 10 GeV/c) 
show the $\rho N$ cross section extracted from
the reaction $\gamma +d \to {\rho}^0 +d$ independently of the
VDM~\cite{Anderson1}. As mentioned before, the results for the 
total cross section
of the $\rho N$ interaction calculated within the framework of the 
quark-model (QM) and the VDM are in fair agreement 
at momenta above 2~GeV/c.
The dotted line in Fig.~\ref{lk5} shows the interpolation
between the low and high energy parts of the total $\rho p$ cross
section which we will adopt furtheron in Eq. (\ref{optic}).

In order to compute the real part of the amplitude we use the dispersion
relation (\ref{dis1}) and perform the substraction at
${\omega}_0=4.46$~GeV since at this energy $Re f({\omega}_0)$ 
was calculated with the  VDM from the $\rho$-meson
photoproduction differential cross section measured by 
DESY-MIT~\cite{DESY} and the Daresbury group~\cite{Daresbury}. 
We note that the VDM should be valid at the energy of 4.46 GeV such 
that the subtraction point is no hidden parameter in our approach.
Since the resonance model is also
valid below the $\rho N$ threshold, the $\rho $
contribution in this case is calculated as an integral 
over the available invariant mass of the $2\pi$ system. 

\begin{figure}[h]
\psfig{file=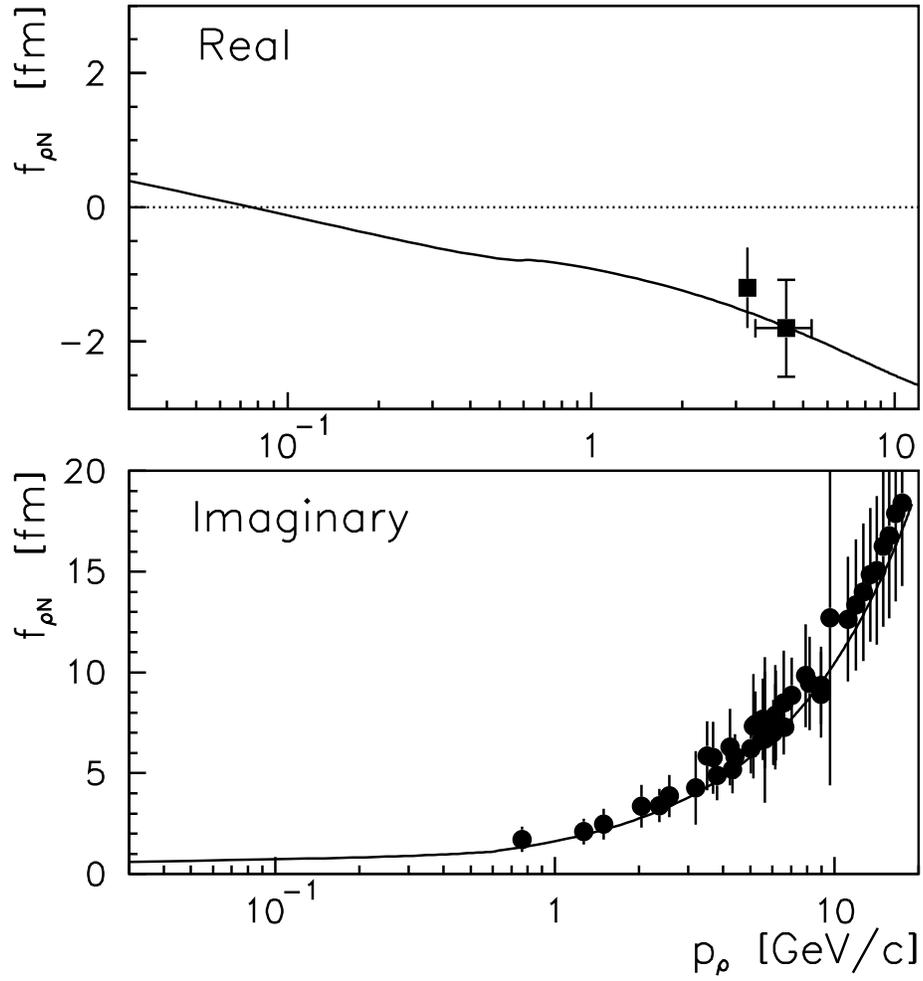,width=15cm}
\caption[]{\label{lk1} Real and imaginary part of the $\rho N$
scattering amplitude in free space from the calculations (solid lines).
The circles and squares are the results evaluated from the 
experimental data for  $\rho$-meson photoproduction.}
\end{figure}

\begin{figure}[h]
\psfig{file=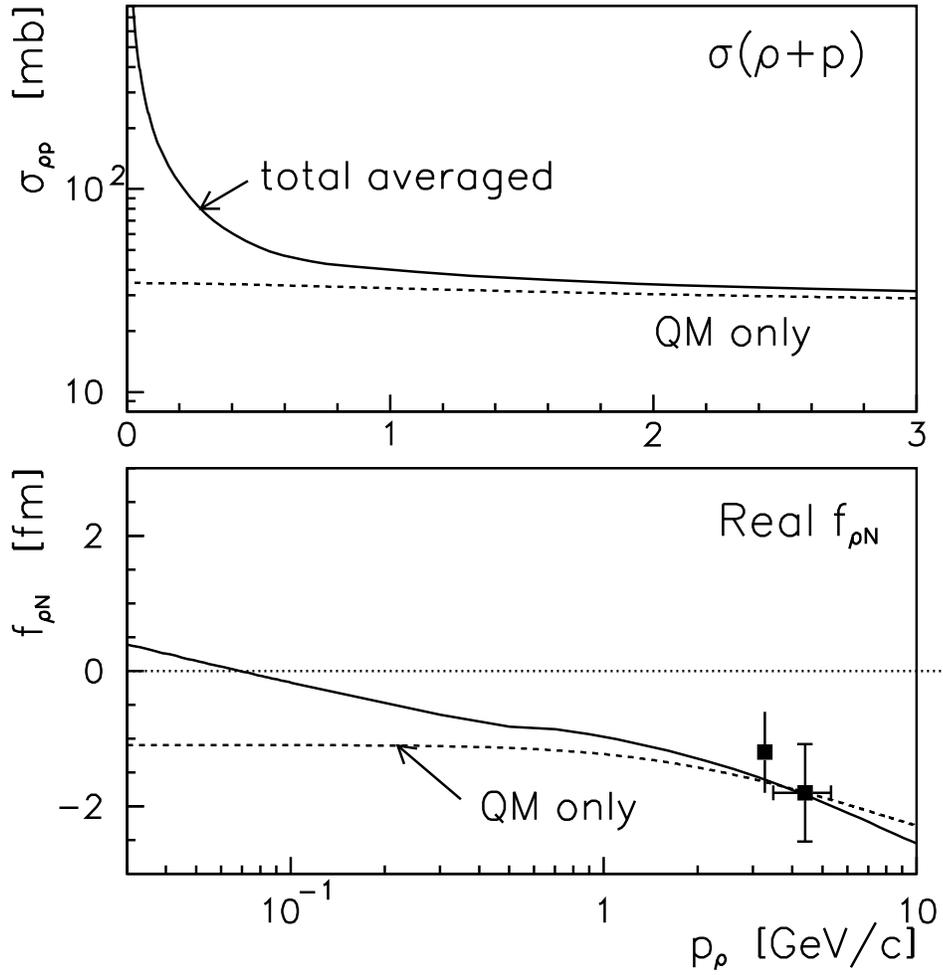,width=15cm}
\caption[]{\label{lk12} The total $\rho p$ cross section and the 
real part of the $\rho N$ scattering amplitude in free space
as a function of the momentum $p_\rho$. The solid lines show our  
calculations with the total $\rho N$ cross section while the
dashed lines indicate the results with the cross section from the
quark model (\protect\ref{QCD1}), only. 
The  squares are evaluated from the experimental 
data for $\rho$-meson photoproduction.}
\end{figure}

The real and imaginary parts of the $f_{\rho N}$ amplitude
- calculated with the total $\rho N$ cross section averaged over 
the $\rho $-meson spectral function in free
space~(\ref{spef}) - are shown in Fig.~\ref{lk1}. The imaginary
part corresponds to the total $\rho p$ cross section from
Fig.~\ref{lk5} using the dashed line as an interpolation. 
Whereas the real part $Re f_{\rho N}$ is negative at high momentum,
it changes the sign at $p_\rho \approx$ 100 MeV/c.
This behaviour of the real part for the $f_{\rho N}$ 
amplitude has its origin
in the resonance contribution at low energies.  To
illustrate this point we present in Fig.~\ref{lk12}  
$Ref_{\rho N}$ neglecting the part of the $\rho N$ cross section
from the resonance model. In fact, in this case the real
part does not change its sign and remains negative over the whole
momentum range.

\begin{figure}[h]
\psfig{file=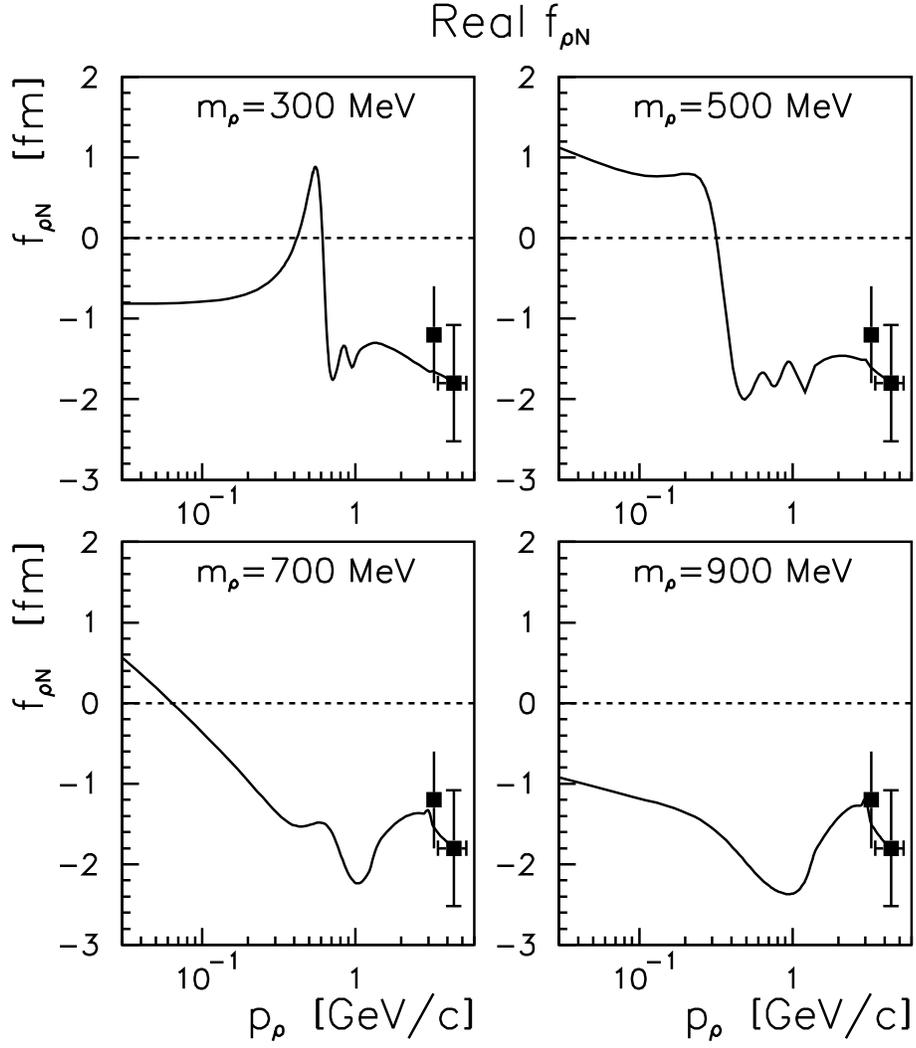,width=15cm}
\caption[]{\label{lk16} The real part of the $\rho N$ scattering 
amplitude in free space calculated for different
masses of the $\rho $-meson. The  squares are evaluated 
from the experimental data for $\rho$-meson photoproduction.}
\end{figure}

Furthermore, Fig.~\ref{lk16} shows the real part of the 
forward $\rho N$ scattering amplitude calculated for the 
different masses of the two-pion system $m$ coupled to the $\rho $-meson.
We find that $Ref_{\rho N}$ substantially depends on the
$\rho $-meson mass as well as on momentum. Note that the dip for 
$p_{\rho}\simeq 1$~GeV/c is due to our interpolation between 
the low and the high energy regions and thus an artefact of our 
model which should be discarded.
We also have to mention that the magnitude of the $\rho $-meson momentum,
where the real part of the forward $\rho N$ scattering
amplitude changes its sign, depends on the prescription
for the transition between the resonance and high
energy part of the total cross section,
which is actually model dependent,  and estimate this
uncertainty as $\delta p_{\rho}=\pm 30$~MeV/c.

\section{Mass shift and broadening of the $\rho$-meson
in the nuclear medium}

\begin{figure}
\psfig{file=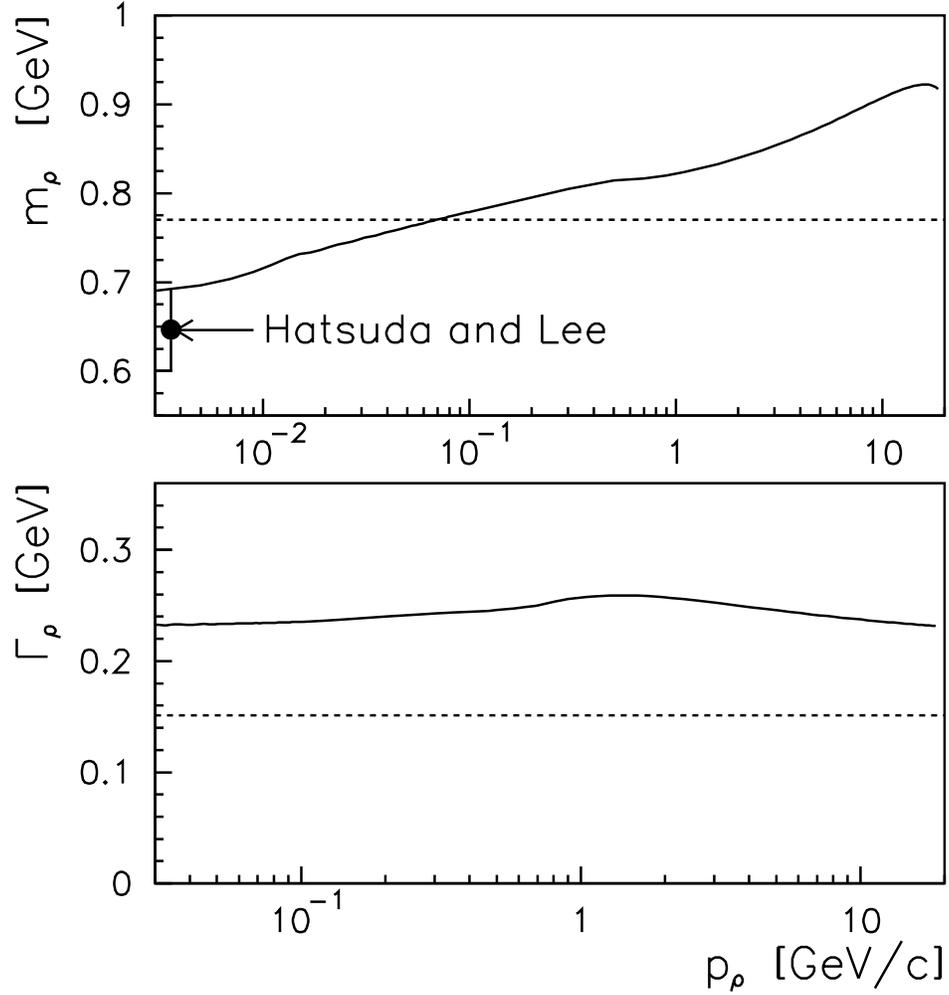,width=15cm}
\caption[]{\label{lk2} Mass and width of the $\rho$-meson at
${\rho}_N = 0.16$~fm$^{-3}$ according to our model in the low density
approximation.  The full dot
corresponds to the result from Hatsuda~\protect\cite{Hatsuda1}
within the QCD sumrule approach.}
\end{figure}

In the low density approximation one now can express the
correction of the $\rho$-meson mass and width at finite nuclear
density $\rho_N$ through $f_{\rho N}$ using 
Eqs.~(\ref{rmassa}) and (\ref{rwidtha}).
We show the corresponding results for the 
mass and width of the $\rho$-meson
in Fig.~\ref{lk2} calculated at saturation density $\rho_0
\approx$ 0.16 $fm^{-3}$ with the averaged $\rho N $
cross section from the resonance model as described above. 
The upper part of Fig.~\ref{lk2} also shows the recent
result from Hatsuda~\cite{Hatsuda1} calculated within the QCD sumrule
approach at $\omega = m_\rho$ by the full dot. Our result
for $\delta m$ in
the low density approximation is in agreement with this calculation
for $p_\rho \approx$ 0. Note that the QCD sumrule analysis 
is also limited to low nuclear densities due to the unknown 
behaviour of the quark 4-point
condensates in the medium.
At high momenta $p \geq $ 2 GeV/c our result is also in qualitative 
agreement with the VDM predictions for the
mass shift as found by Eletsky and Ioffe~\cite{Eletsky}. 
In their case the $\rho$ mass shift increases from 60~MeV at 
$p_\rho =2$~GeV/c to 90~MeV at 7~GeV/c, whereas our analysis 
gives a shift of $\simeq$70~MeV at 2~GeV/c and $\simeq$120~MeV at
7~GeV/c. The deviations
with the results from~\cite{Eletsky}  are due to the use of the 
$\rho$-photoproduction amplitudes (\ref{vector}) instead of the Compton
scattering amplitude (\ref{coupl}) used by the latter authors.

Thus, if the low density approximation could be extrapolated up to
$\rho_0$, our model would
demonstrate that the repulsive interaction of the $\rho$-meson at high
momenta might be consistent with an attractive $\rho N$ interaction 
at low energies.   
However, as it was shown  Ref.~\cite{Mosel}, the effect 
of the finite $\rho$ width  can
essentially influence the predictions of QCD sum 
rules~\cite{Hatsuda,Asakawa}.
Furthermore, recent microscopic calculations of the $\rho$-spectral
function at low energies do not show any substantial attraction
for slow $\rho$-mesons at normal nuclear density~\cite{Klingl,Peters}.

It is thus important to check if the results obtained within the low
density approximation are still valid at normal nuclear density $\rho_0$.
Up to now in calculating the mass shift and 
broadening of the width for the
$\rho$-meson (shown in Fig.~\ref{lk2}) we have used the vacuum $\rho N$
scattering amplitude with resonance
contributions which, however,  might be different in nuclear matter.
An indication for a possible strong medium-modification 
of baryonic resonances is the experimental observation
that the total photoabsorption cross section in nuclei for
$A \ge 12$ does not show any resonant structure except the 
$\Delta (1232)$ isobar~\cite{Frascati,Frascati2,Mainz}. 
As was shown in Ref.~\cite{Kondratyuk,Boffi}
this might be explained by a strong in-medium broadering of 
the $D_{13}(1520)$ resonance (${\Gamma}_{med} \simeq 300$~MeV
compared to the vacuum width $\Gamma_R\simeq 120$~MeV),
but not by conventional medium effects like Fermi motion or
Pauli blocking. While a collision broadering of this order
is hard to justify, it might arise from the strong coupling of
the $D_{13}$ resonance to the $\rho N$ channel and a
medium modification of the $\rho $-meson~\cite{Martin}.
Furthermore, in Ref.~\cite{Peters} it has been shown that such
a mechanism can lead to an in-medium
width of the $D_{13}(1520)$ resonance of about 350~MeV.

To estimate how the latter effect influences our results
we have performed also calculations for the in-medium $\rho N$ total 
cross section by  assuming  that at normal nuclear density the
widths of all resonances -  coupled to the $\rho N$ channel -  are
twice as in the vacuum, but their $\rho N$ branching ratios
stay the same. This
in-medium $\rho N$ cross section then was used in the dispersion
relation for the calculation of the in-medium real part  of the $\rho N$
forward scattering amplitude.
 
\begin{figure}
\psfig{file=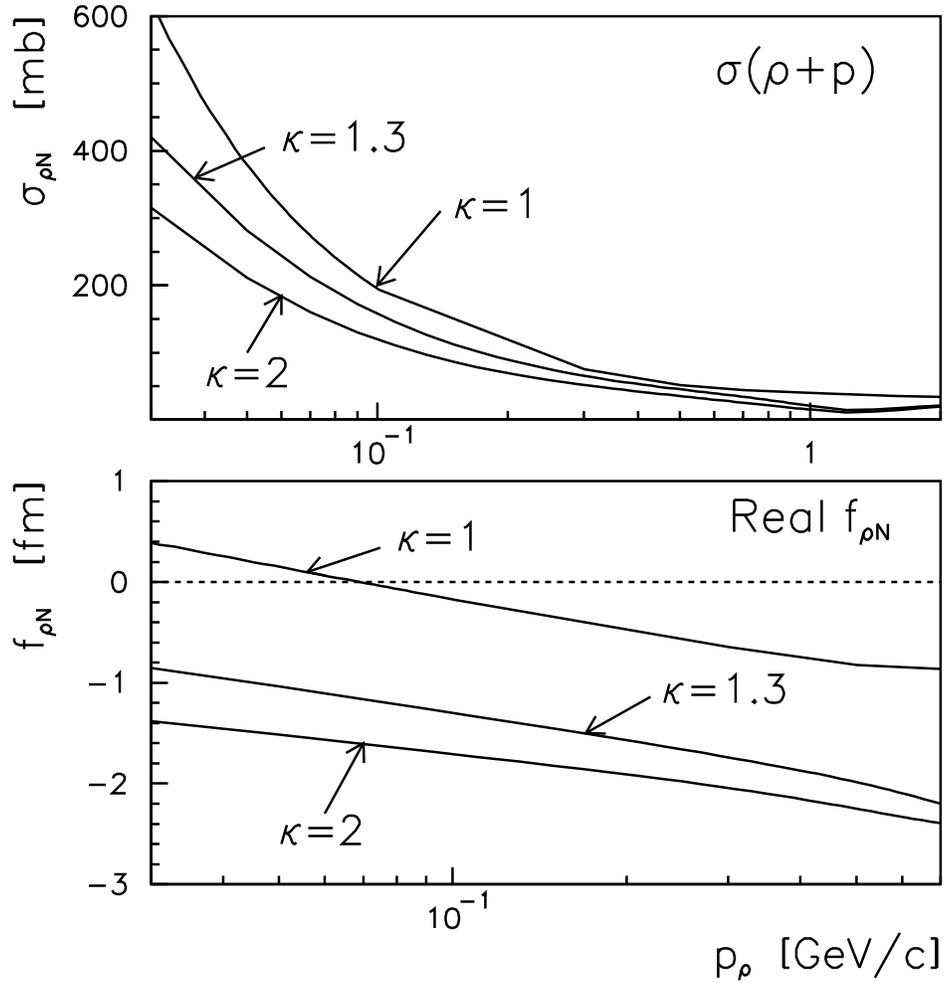,width=15cm}
\caption[]{\label{lk13} The total $\rho N$ cross section and the 
real part of the $\rho N$ scattering amplitude calculated
for different widths of the baryonic resonances. 
The factor $\kappa $ indicates the ratio of the in-medium resonance
width to its vacuum width.}
\end{figure}

The results of these calculations are shown in Fig.~\ref{lk13}
for the total $\rho p$ cross
section and the real part of the forward scattering amplitude
calculated with different widths of all baryonic resonances. 
The factor $\kappa $ in Fig.~\ref{lk13}
stands for the ratio of the in-medium width of the baryonic 
resonance to its value in free space.
We see that the higher order resonance broadening effects have a strong 
influence on the
$\rho N$ scattering amplitude in the resonance region. The total cross
section becomes smoother at low energy and as a result
the real part of $f_{\rho N}(0)$ does not change sign anymore with
decreasing  $p_\rho$. At small $p_\rho$ it remains negative,
which means that the main medium effect for the 
$\rho$-meson  at normal nuclear
density is the collisional broadening; the 
mass shift is slightly repulsive.
This result is in qualitative agreement with recent calculations
of the $\rho$-meson spectral function at low energies from 
Refs.~\cite{Klingl,Rapp,Peters}.  However, a note of caution has to be 
added here because the predictions within the resonance model are only of
qualitative nature. The properties of nuclear resonances as well as their
branching to the $\rho$ channel in the dense medium are unknown so far.
Furthermore, the influence of exchange-current corrections at high density
on the $\rho$ meson properties should be considered as well.

\section{Conclusions}

In summary, we have calculated the momentum dependence of the
$\rho$-meson selfenergy (or in-medium properties) based on the
dispersion relation~(\ref{dis1}) for the $\rho N$ scattering amplitude at
finite (but small) nuclear density. The imaginary part of $f(\omega)$ is
calculated via the optical theorem~(\ref{optic}) while the total $\rho N$
cross section is obtained within the VDM at high energy and
within the resonance model at low energy. The 
scattering amplitude $f(\omega)$
thus is entirely based on experimental data. Our numerical results
indicate a sizeable broadening of the $\rho$-meson width in the
medium especially for low relative momenta $p_\rho$.
In the low density approximation the real part of its selfenergy is 
found to be attractive for $p_\rho \leq 100$~MeV/c and to change its sign,
becoming repulsive at higher momenta in line with the
prediction by Eletsky and Joffe~\cite{Eletsky}.   
Extrapolating the low density approximation to
nuclear saturation density we obtain
a dropping of the $\rho$-mass at $p_\rho \approx$ 0 in line with the
QCD sumrule analysis of Hatsuda~\cite{Hatsuda1}. Thus our dispersion
approach demonstrates that the results of Refs. \cite{Eletsky} and Ref.
\cite{Hatsuda1} do not contradict each other due to the rather strong
momentum dependence of the $\rho N$ scattering amplitude. 

However, the resonance part of the $\rho N$ scattering
amplitude is also influenced by the nuclear medium at
saturation density such that the moderate  attraction at $p_\rho \simeq$0
changes to a small repulsion. This behaviour of
the $\rho$-meson selfenergy is in qualitative agreement with
the recent microscopic calculations of Rapp, Chanfray and 
Wambach~\cite{Rapp,Wambach} as well as Klingl et al.~\cite{Klingl} 
for the $\rho$ spectral function. In the latter case the $\rho$-meson
essentially broadens significantly in the dense medium which implies
lifetimes of the $\rho$-meson less than 1 fm/c already at 
density $\rho_0$; in simple words:
according to our dispersion approach
the $\rho$-meson 'melts' at high nuclear density and 
does not 'drop in mass'
as suggested by the scaling hypothesis of Ref. \cite{Brown}. 

We, finally, note that the explicit
momentum dependence of the $\rho$-meson selfenergy is also an
important issue that has to be incorporated e.g. in transport
theories that attempt to extract information on the $\rho$
spectral function in comparison to experimental dilepton data.

\vspace{1cm}
The authors acknowledge many helpful discussions with K.
Boreskov, E.L. Bratkovskaya, B.L. Ioffe, U. Mosel and Yu. Simonov
throughout this study.

\newpage
\begin{table*}[h]
\begin{center}
\caption{\label{ta1} Properties of baryonic resonances
coupled to the $\rho$-meson; $l$ denotes the angular momentum 
of the $\rho$ meson while the stars indicate the confidence level.
In our calculations we discard resonances with only one star C.L.}
\vspace{0.6cm}
\begin{tabular}{|l|c|c|c|c|c|c|}
\hline
Resonance & $J_R$ & $M_R$ (MeV) & ${\Gamma}_R $ (MeV) &
$l$ & ${\to}N{\rho}$ (\%) & C.L.  \\
\hline
$ S_{11}(1650) $  & 1/2 & 1659 & 173 & 0 & 3 & $\ast \ast \ast \ast $\\
$ S_{11}(2090) $  & 1/2 & 1928 & 414 & 0 & 49 & $ \ast $ \\
$ D_{13}(1520) $  & 3/2 & 1524 & 124 & 0 & 21 & $ \ast \ast \ast \ast $ \\
$ D_{13}(1700) $  & 3/2 & 1737 & 249 & 0 & 13 & $ \ast \ast \ast $\\
$ D_{13}(2080) $  & 3/2 & 1804 & 447 & 0 & 26 & $ \ast \ast$  \\
$ G_{17}(2190) $  & 7/2 & 2127 & 547 & 2 & 29 & $ \ast \ast \ast \ast $\\
$ P_{11}(1710) $  & 1/2 & 1717 & 478 & 1 & 3 & $ \ast \ast \ast $ \\
$ P_{11}(2100) $  & 1/2 & 1885 & 113 & 1 & 27 & $ \ast $\\
$ P_{13}(1720) $  & 3/2 & 1717 & 383 & 1 & 87 & $ \ast \ast \ast \ast $ \\
$ P_{13}       $  & 3/2 & 1879 & 498 & 1 & 44 & \\
$ F_{15}(1680) $  & 5/2 & 1684 & 139 & 1 & 5 & $ \ast \ast \ast \ast $ \\
$ F_{15}(1680) $  & 5/2 & 1684 & 139 & 3 & 2 & $ \ast \ast \ast \ast $ \\
$ F_{15}(2000) $  & 5/2 & 1903 & 494 & 1 & 60 & $ \ast \ast$ \\
$ F_{15}(2000) $  & 5/2 & 1903 & 494 & 3 & 15 & $ \ast \ast$ \\
$ S_{31}(1620) $  & 1/2 & 1672 & 154 & 0 & 29 & $ \ast \ast \ast \ast $\\
$ S_{31}(1900) $  & 1/2 & 1920 & 263 & 0 & 38 & $ \ast \ast \ast $\\
$ D_{33}(1700) $  & 3/2 & 1762 & 599 & 0 & 8 & $ \ast \ast \ast \ast $\\
$ D_{33}(1940) $  & 3/2 & 2057 & 460 & 0 & 35 & $ \ast $ \\
$ P_{31}(1910) $  & 1/2 & 1882 & 239 & 1 & 10 & $ \ast \ast \ast \ast $\\
$ F_{35}(1905) $  & 5/2 & 1881 & 327 & 1 & 86 & $ \ast \ast \ast \ast $\\
$ F_{35}       $  & 5/2 & 1752 & 251 & 1 & 22 & \\  
\hline
\end{tabular}
\end{center}
\end{table*}

\clearpage

\newpage

\newpage

\newpage

\newpage

\newpage

\newpage

\newpage

\end{document}